\documentclass[twocolumn,showpacs,preprintnumbers,amsmath,amssymb]{revtex4}

\usepackage{graphicx}
\usepackage{dcolumn}
\usepackage{bm}
\usepackage{amsmath}

\begin{document}
\title{Synchronization on community networks}
\author{Tao Zhou$^{1,2}$}
\author{Ming Zhao$^{1}$}
\author{Guanrong Chen$^{2}$}
\author{Gang Yan$^{3}$}
\author{Bing-Hong Wang$^{1}$}
\affiliation{%
$^{1}$Department of Modern Physics and Nonlinear Science Center,
University of Science and Technology of China, Anhui Hefei 230026,
PR China\\
$^{2}$Department of Electronic Engineering, City University of
Hong Kong, Hong Kong SAR, PR China \\
$^{3}$Department of Electronic Science and Technology, University
of Science and Technology of China, Hefei Anhui, 230026, PR China
}%

\date{\today}

\begin{abstract}
In this Letter, we propose a growing network model that can
generate scale-free networks with a tunable community strength.
The community strength, $C$, is directly measured by the ratio of
the number of external edges to internal ones; a smaller $C$
corresponds to a stronger community structure. According to the
criterion obtained based on the master stability function, we show
that the synchronizability of a community network is significantly
weaker than that of the original Barab\'asi-Albert network.
Interestingly, we found an unreported linear relationship between
the smallest nonzero eigenvalue and the community strength, which
can be analytically obtained by using the combinatorial matrix
theory. Furthermore, we investigated the Kuramoto model and found
an abnormal region ($C\leq 0.002$), in which the network has even
worse synchronizability than the uncoupled case ($C=0$). On the
other hand, the community effect will vanish when $C$ exceeds 0.1.
Between these two extreme regions, a strong community structure
will hinder global synchronization.
\end{abstract}


\maketitle

Synchronization is observed in a variety of natural, social,
physical and biological systems, and has found applications in a
variety of fields \cite{Strogatz2003}. The large number of
networks of coupled dynamical systems that exhibit synchronized
states are subjects of great interest. In the early stage, these
studies are restricted to either the regular networks
\cite{Regular}, or the random ones \cite{Random}. Recently,
inspired by the new discovery of several common characteristics of
real networks, the majority of the studies about network
synchronization focus on networks with complex topologies. The
effects of average distance \cite{AverageDistance}, heterogeneity
\cite{Heterogeneity}, clustering \cite{Clustering}, and weight
distribution \cite{Weight} on network synchronizability have been
extensively investigated.

\begin{figure}
\scalebox{0.40}[0.45]{\includegraphics{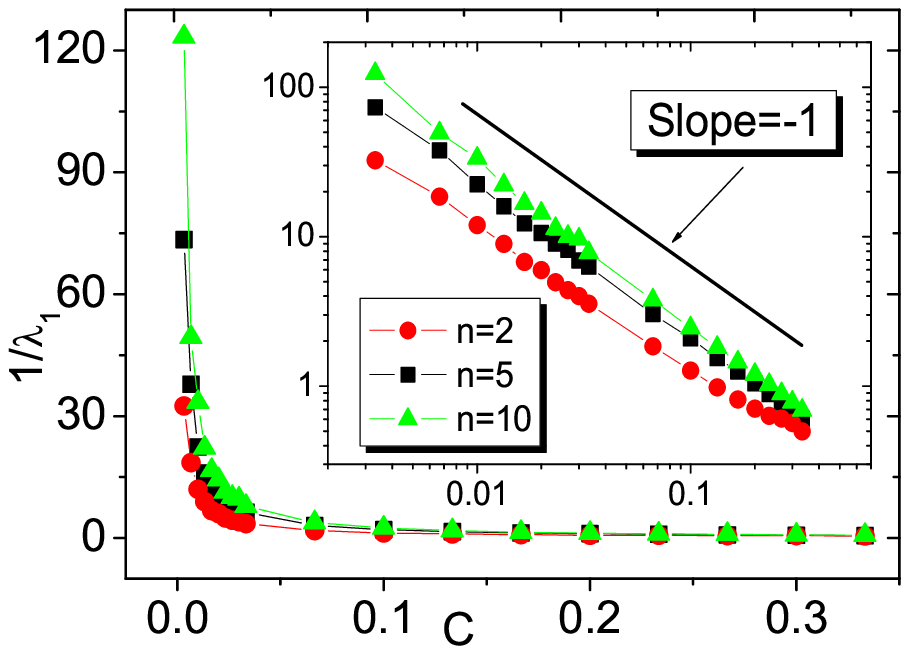}}
\scalebox{0.40}[0.45]{\includegraphics{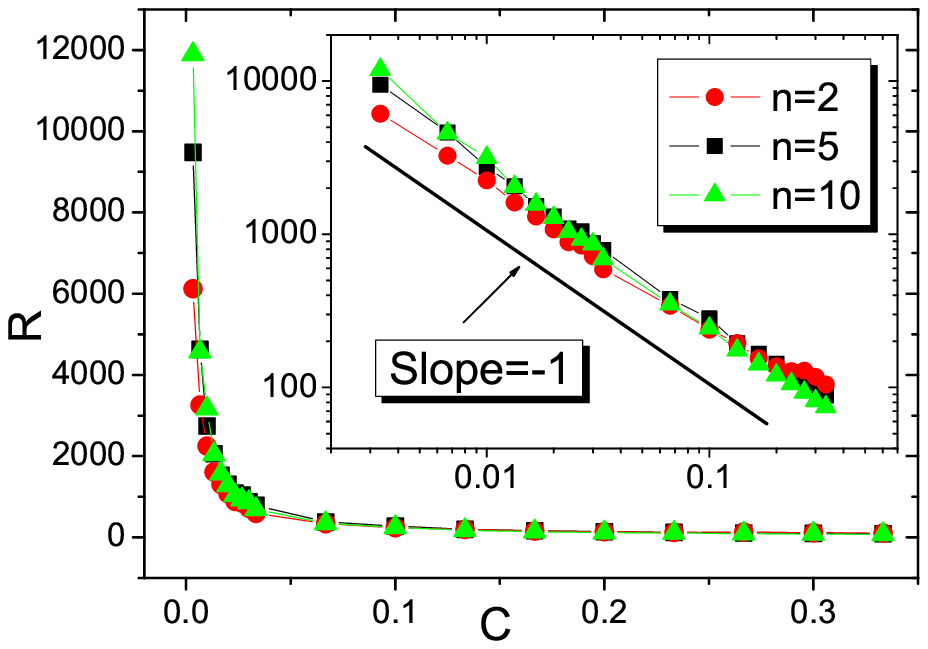}}\caption{(Color
online) The inverse of the smallest nonzero eigenvalue
$1/\lambda_1$ (left) and the eigenratio $R$ (right) vs $C$. The
red circles, black squares, and green triangles represent the
cases of $n=2$, $n=5$, and $n=10$, respectively. As shown in the
insets, the same data can be well fitted by a straight line with
slope $\approx -1$ in the log-log plot, indicating the relation
$\frac{1}{\lambda_1}\sim C^{-1}$. All the data are obtained as the
average over 10 realizations, and for each realization, the
network parameters $N=5000$ and $m=3$ are fixed.}
\end{figure}

Besides the small-world and scale-free properties, it has been
demonstrated that many real networks have the so-called
\emph{community structure} \cite{Community}. Qualitatively, a
community is defined as a subset of nodes within a network such
that connections between the nodes therein are denser than that
with the rest of the network \cite{Radicchi2004}. Very recently,
by applying the epidemiological models on community networks, it
was found that the network epidemic dynamics are highly affected
by the community structure \cite{Epidemic}. To date, however, the
issue of synchronization on community networks has not been fully
investigated. Based on a toy network model with a tunable
\emph{community strength}, in this Letter we intend to provide a
first analysis on how community structure affects the network
synchronizability.

Our model starts from $n$ community cores, each core contains
$m_0$ fully connected nodes. Initially, there are no connections
among different community cores. At each time step, for each
community core, one node is added. Thus, there are in total $n$
new nodes being added in one time step. Each node will attach $m$
edges to existing nodes within the same community core, and
simultaneously $m'$ edges to existing nodes outside this community
core. The former are \emph{internal edges}, and the latter are
\emph{external edges}. Note that, the $m$ and $m'$ are not
necessary to be integers, for example, to generate 2.7 edges can
be implemented as follows: Firstly, generate 2 edges, and then
generate the third one with probability 0.7. Similar to the
evolutionary mechanism of Barab\'asi-Albert (BA) networks
\cite{Barabasi1999}, we assume the probability of choosing an
existing node $i$ to connect to is proportional to $i$'s degree
$k_i$. Each community core will finally become a single community
of size $N_c$, and the network size $N=nN_c$. By using the
rate-equation approach \cite{Krapivsky2000}, one can easily obtain
the degree distribution of the whole network, $p(k)\sim k^{-3}$.
For simplicity, we directly use the ratio of external edges to
internal ones, $C=\frac{m'}{m}$, to measure the strength of the
community structure. Clearly, a smaller $C$ corresponds to sparser
external edges thus a \emph{stronger} community structure.

Next, we investigate how the community strength $C$ affects the
network synchronizability. Consider $N$ identical dynamical
systems (oscillators) with the same output function, which are the
nodes of a network and coupled linearly and symmetrically with
neighbors through edges. The coupling fashion ensures the
synchronization manifold be an invariant manifold, and the
dynamics can be locally linearized near the synchronous state. The
state of the $i$th oscillator is described by $\textbf{x}^i$, and
the set of equations of motion governing the dynamics of the $N$
coupled oscillators is
\begin{equation}
\dot{\textbf{x}}^i=\textbf{F}(\textbf{x}^i)-\sigma\sum_{j=1}^NG_{ij}\textbf{H}(\textbf{x}^j),
\end{equation}
where $\dot{\textbf{x}}^i=\textbf{F}(\textbf{x}^i)$ governs the
dynamics of the $i$th individual oscillator,
$\textbf{H}(\textbf{x}^j)$ is the output function, $\sigma$ is the
coupling strength, and the $N\times N$ Laplacian $\textbf{G}$ is
given by
\begin{equation}
    G_{ij}=\left\{
    \begin{array}{cc}
    k_i   &\mbox{for $i=j$}\\
     -1    &\mbox{for $j\in\Lambda_i$}\\
     0    &\mbox{otherwise}.
    \end{array}
    \right.
\end{equation}

Being positive semidefinite, all the eigenvalues of $\textbf{G}$
are nonnegative reals and the smallest eigenvalue $\lambda_0$ is
always a single zero, for the rows of $\textbf{G}$ have zero sums.
Thus, the eigenvalues can be ranked as
0=$\lambda_0<\lambda_1\leq\cdots\leq\lambda_{N-1}$. If the
synchronized region is left-unbounded, according to the Wang-Chen
(WC) criterion \cite{WangChen}, the network synchronizability can
be measured by the inverse of the smallest nonzero eigenvalue
$1/\lambda_1$: the smaller the $1/\lambda_1$, the better
synchronizability, and vice versa. We show the numerical results
about the relation between $1/\lambda_1$ and $C$ in the left plot
of Fig. 1. Clearly, the community structure will hinder the global
synchronization. More interestingly and significantly, we found an
unreported linear relationship, $\lambda_1\propto C$. This seems a
universal law for community networks if the community structure is
sufficiently strong (i.e. $C$ is sufficiently small).

\begin{figure}
\scalebox{0.85}[0.85]{\includegraphics{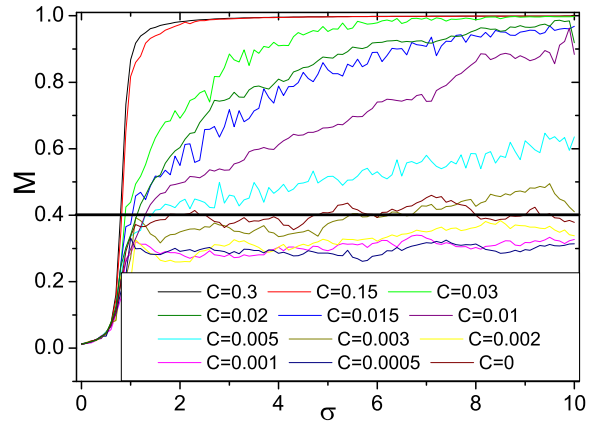}} \caption{(Color
online) Order parameter $\mathbb{M}$ vs coupling strength $\sigma$
for different values of the community strength $C$. The solid line
represents $\mathbb{M}=0.403$. All the data are obtained as the
average over 100 realizations. For each realization, the network
parameters $N=5000$, $n=5$, $N_c=1000$ and $m=3$ are fixed.}
\end{figure}

Denote by $G_i$ the $N_c\times N_c$ Laplacian of the $i$th
community, $H_i$ the submatrix of $G$ consisting of all the rows
and columns corresponding to the nodes of the $i$th community.
Note that $H_i$ and $G_i$ are different only in diagonal elements,
and $H_i$'s smallest eigenvalue $\lambda^H_{i0}$ is positive. By
using the combinatorial matrix theory, one can first prove that
the smallest nonzero eigenvalue $\lambda_1$ of $G$ is equal to the
minimal one of all the smallest eigenvalues $\lambda^H_{i0}$
($i=0,1,\cdots,n $): $\lambda_1=\texttt{min}_{1\leq i \leq
n}\lambda^H_{i0}$. And, then, one is able to prove that for each
matrix $H_i$, $\lambda^H_{i0}$ is approximately linearly
correlated with the community strength $C$. The strict and full
proof is fairly complicated and is omitted here.

If the synchronized region is finite, according to the
Pecora-Carroll-Barahona (PCB) criterion
\cite{PecoraCarrollBarahona}, the network synchronizability can be
measured by the eigenratio $R=\lambda_{N-1}/\lambda_1$: the
smaller it is, the better synchronizability will be, and vice
versa. We also have checked that the maximal eigenvalue is not
sensitive to the change of the community strength (it slowly
diminishes as $C$ decreases), thus both the WC and PCB criteria
give qualitatively the same result (see the right plot of Fig. 1
for the case of the PCB criterion).

\begin{figure}
\scalebox{0.85}[0.85]{\includegraphics{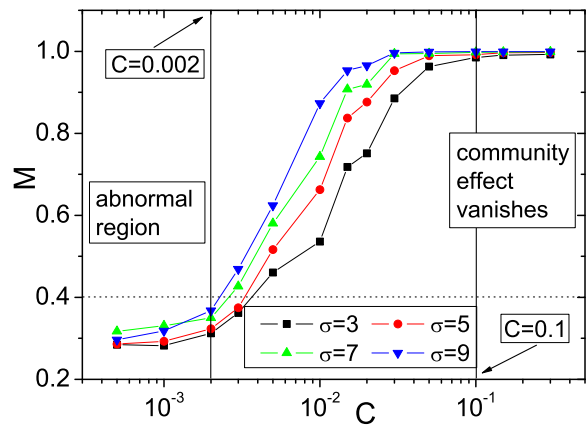}} \caption{(Color
online) Order parameter $\mathbb{M}$ vs community strength $C$ for
different values of the coupling strength $\sigma$. The dash line
represents $\mathbb{M}=0.403$. All the data are obtained by the
average over 100 realizations. For each realization, the network
parameters $N=5000$, $n=5$, $N_c=1000$ and $m=3$ are fixed.}
\end{figure}

Hereinafter, we investigate a network of nonidentical Kuramoto
oscillators \cite{Kuramoto}, obeying the coupled differential
equations
\begin{equation}
\frac{d\phi_i}{dt}=\omega_i+\frac{\sigma}{k_i}\sum_ja_{ij}\texttt{sin}(\phi_i-\phi_j),
\end{equation}
where $0\leq \phi_i<2\pi$ are phase variables, $\omega_i$ are
intrinsic frequencies, $i=1,2,\cdots,N$, $\sigma$ is the coupling
strength, and $[a_{ij}]$ is the adjacency matrix ($a_{ij}=1$ iff
nodes $i$ and $j$ are connected). Initially, $\phi_i$ and
$\omega_i$ are randomly and uniformly distributed in the intervals
$[0,2\pi)$ and $[-0.5,0.5]$, respectively. The numerical results
are obtained by integrating Eqs. (3) using the Runge-Kutta method
with step size 0.01. To characterize the synchronized states, we
use the order parameter
\begin{equation}
\mathbb{M}=\left\{\left|\frac{1}{N}\sum^N_{j=1}e^{i\phi_j}\right|\right\},
\end{equation}
where $\{\cdot\}$ signifies the time averaging. The order
parameters are averaged over $10^4$ time steps, excluding the
former 5000 time steps, to allow for relaxation to a steady state.
Clearly, $\mathbb{M}$ is of order $1/\sqrt{N}$ if the oscillators
are completely uncorrelated, and will approach 1 if they are all
in the same phase.

Fig. 2 reports the simulation results for different community
strength $C$. Whatever the value of $C$, the parameter
$\mathbb{M}$ increases sharply after a critical point
$\sigma_c=0.6$. This point is just the same as the critical point
at which synchronized behavior emerges in each separate community
(BA network of size 1000 and with average degree 6). For all the
cases with $C\geq 0.003$, a strong community structure (i.e. a
smaller $C$) hinders global synchronization. It is difficult to
harmonize different communities based only on a very few external
edges. The community effect becomes lower as $C$ increases. For
sufficiently large $C$ (see the cases of $C=0.15$ and $C=0.30$ in
Fig. 2), the network synchronized behavior is almost the same as
that of the original BA network; that is, community effect
vanishes.

Consider a network consisting of $n$ uncoupled communities
($C=0$), and each community itself can approach a nearly
completely synchronized state. The order parameter $\mathbb{M}$ of
the whole network is
\begin{equation}
\mathbb{M}(n)=\frac{1}{n(2\pi)^n}\int_0^{2\pi}\texttt{d}\phi_1\int_0^{2\pi}\texttt{d}\phi_2\cdots
\int_0^{2\pi}\texttt{d}\phi_n\chi,
\end{equation}
where
$\chi=\sqrt{(\sum_i\texttt{sin}\phi_i)^2+(\sum_i\texttt{cos}\phi_i)^2}$.
The numerical result $\mathbb{M}(5)\approx 0.403$ is represented
by a horizontal line in Fig. 2. The corresponding simulation with
$C=0$ shows that $\mathbb{M}$ slightly fluctuates around 0.403 for
sufficiently large $\sigma$, which is in accordance with the above
numerical result. Interestingly, we found an abnormal region
$C\leq 0.002$, in which the networks have even worse
synchronizability than the uncoupled (unconnected) case. Within
this region, each separate community can not get harmonized with
other communities through its few external edges. On the contrary,
the input signals containing by these edges disturb the
synchronizing process of this community.

Fig. 3 shows the order parameter $\mathbb{M}$ vs $C$ for different
$\sigma$. As mentioned above, when $C\leq 0.002$, $\mathbb{M}$ can
not reach the dash line ($\mathbb{M}$=0.403). We have also checked
that even for very large $\sigma$, $\mathbb{M}$ is always smaller
than 0.403 if $C\leq 0.002$. The distinct difference between the
original BA networks and the present community networks vanishes
when the density of external edges exceeds 0.1. The division of
those three regions is qualitative, and the borderlines between
neighboring regions can not be exactly determined. However, it
provides a clearer picture about the effect of the community
structure.

\begin{figure}
\scalebox{0.39}[0.45]{\includegraphics{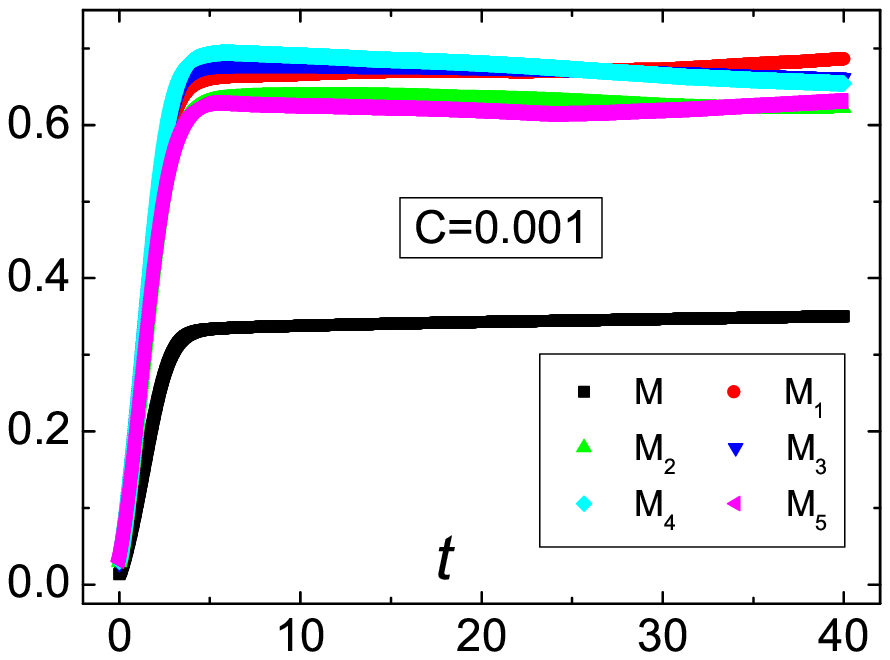}}
\scalebox{0.39}[0.45]{\includegraphics{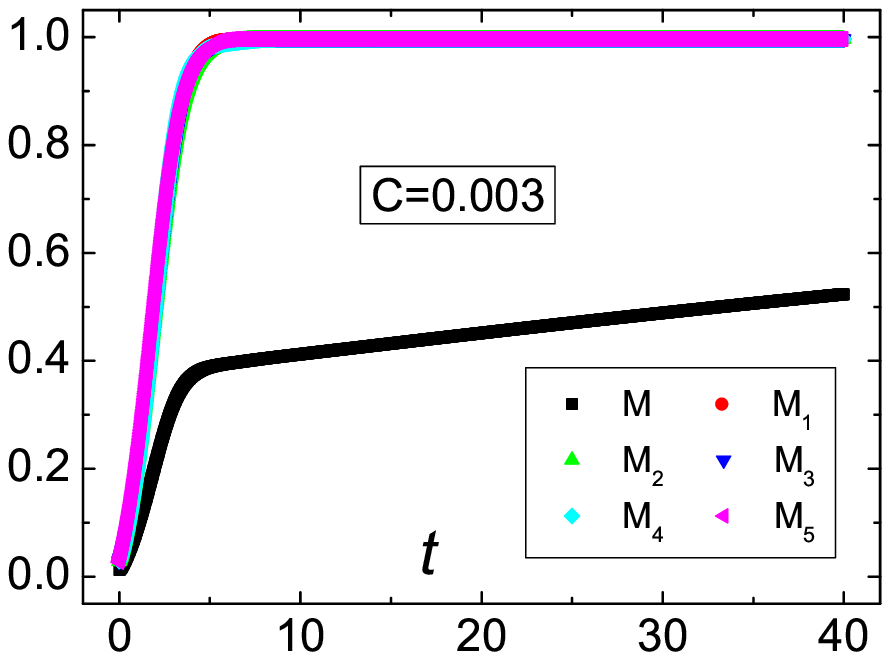}}
\scalebox{0.39}[0.45]{\includegraphics{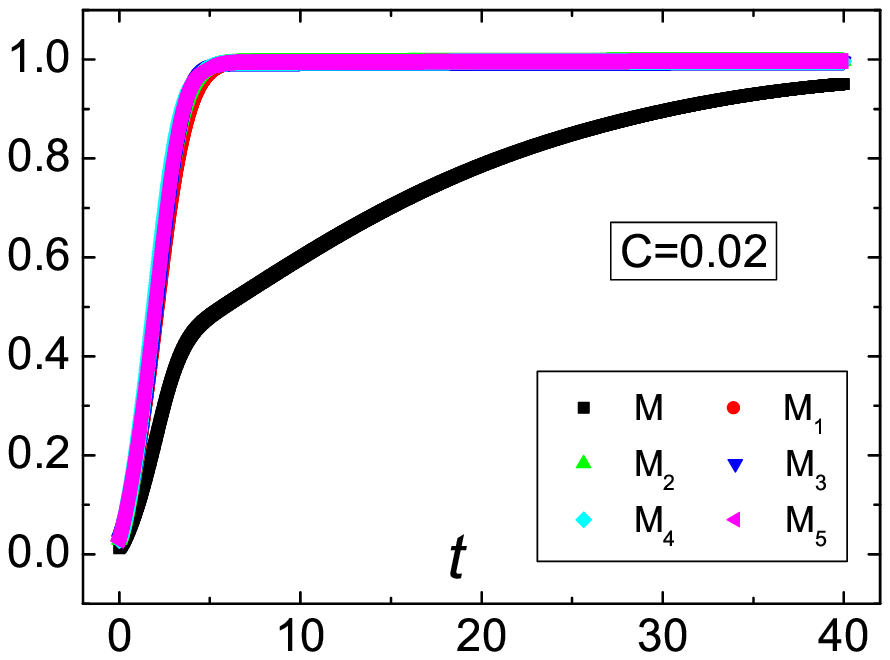}}
\scalebox{0.39}[0.45]{\includegraphics{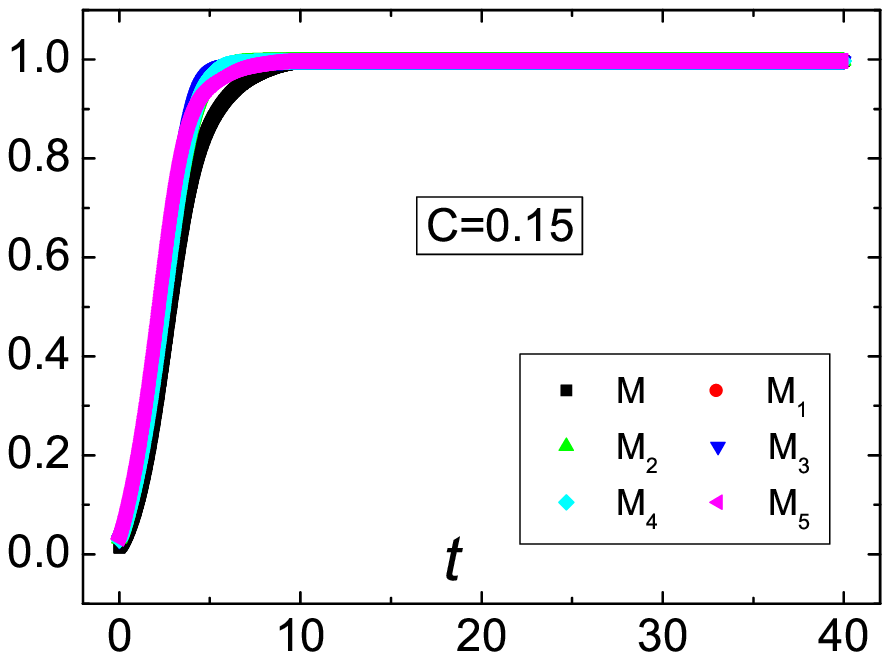}} \caption{(Color
online) The order parameters vs time for the whole network and
each community. The network parameters are $N=5000$, $n=5$,
$N_c=1000$ and $m=3$. The coupling strength $\sigma=5$ is fixed.}
\end{figure}

To further understand the underlying mechanism of synchronization
on community networks, we investigate the partial synchronization
within a separate community. For the $i$th community, the
corresponding order parameter $\mathbb{M}_i$ is defined as
\begin{equation}
\mathbb{M}_i=\left\{\left|\frac{1}{N_c}\sum_je^{i\phi_j}\right|\right\},
\end{equation}
where the sum goes over all the nodes belonging to the $i$th
community. Fig. 4 exhibits the temporal behaviors of order
parameters for the whole network and for each community. The four
plots correspond to the cases of $C=0.001$, $C=0.003$, $C=0.02$,
and $C=0.15$, respectively. In the abnormal region ($C=0.001$),
due to the external disturbance, the order parameter of each
community is remarkably below 1 even in the long time limit. After
$\mathbb{M}_i$ reaches its steady value ($t\approx 4$),
$\mathbb{M}$ also becomes steady, indicating that the external
edges can not harmonize different communities, but only introduce
some noise thus hinder the partial synchronization. When $C\geq
0.003$, a separate community can approach a nearly completely
synchronized state, and can harmonize with other communities;
therefore, $\mathbb{M}$ will continuously increase after
$\mathbb{M}_i$ gets steady. For sufficiently large $C$ (see the
case of $C=0.15$), the whole network can approach the nearly
completely synchronized state almost as quickly as the single
separate community, and the effect of the community structure on
network dynamics (at least for the Kuramoto model) is hardly
observed.

In conclusion, we have proposed propose a scale-free network model
with a tunable community strength and studied its synchronization
phenomenon. We found an unreported linear relationship
$\lambda_1\propto C$, which is the first quantitative formula that
describes the synchronizability of community networks. We have
also checked that the maximal eigenvalue is not sensitive to the
change of the community strength, thus both the Wang-Chen and
Pecora-Carroll-Barahona criteria give qualitatively the same
result: The stronger the community structure, the worse the
synchronizability. Furthermore, we have investigated the Kuramoto
model in community networks. Interestingly, we found an abnormal
region, in which the networks have even worse synchronizability
than the uncoupled case. Due to the complicacy of the scale-free
structure itself, we are unable to give a theoretical and
analytical explanation about this observed phenomenon. An
approximate analytic solution of a similar phenomenon may be
obtained based on a more ideal community network model, where each
community is a complete graph (Very recently, this ideal model has
also been investigated to show the effect of modular number on
network synchronization \cite{Park2006}). Beyond this abnormal
region, analogous to the result from the approach of the master
stability function, increasing the density of external edges will
sharply enhance the network synchronizability. However, when the
density of external edges exceeds 0.1, the synchronized behavior
becomes almost the same as that of the original BA networks and
further enhancement can not be achieved. This result is not only
of theoretical interest, but also significant in practice if one
wants to enhance the synchronizability of community networks by
adding external edges. Finally, we would like to point out that,
although the present model is very simple, it provides a useful
way to have detailed understanding about the effect of the
community structure on network dynamics since the community
strength in the model is adjustable. We believe that this model
can also be applied to the studies on many other network dynamical
processes.

This work was partially supported by the National Natural Science
Foundation of China under Grant Nos. 70471033, 10472116, and
70571074, the Specialized Research Fund for the Doctoral Program
of Higher Education (SRFDP No.20020358009), the Special Research
Founds for Theoretical Physics Frontier Problems under Grant No.
A0524701, and Specialized Program under the Presidential Funds of
the Chinese Academy of Science.

\end{document}